\documentclass[12pt,preprint]{aastex}

\shorttitle{}
\shortauthors{Nesvorn\'y et al.}

\begin{document}

\title{Fast Inversion Method for Determination of Planetary Parameters from Transit 
Timing Variations}

\author{David Nesvorn\'y$^1$ and Cristi\'an Beaug\'e$^2$}
\affil{(1) Department of Space Studies, Southwest Research Institute,
1050 Walnut St., Suite 400, Boulder, Colorado 80302, USA}
\affil{(2) Observatorio Astron\'omico, Universidad Nacional de C\'ordoba, Laprida 854, 
C\'ordoba, X5000BGR, Argentina} 

\begin{abstract}
The Transit Timing Variation (TTV) method relies on monitoring changes in timing of transits
of known exoplanets. Non-transiting planets in the system can be inferred from TTVs by their
gravitational interaction with the transiting planet. The TTV method is sensitive to low-mass
planets that cannot be detected by other means. Here we describe a fast algorithm that can be 
used to determine the mass and orbit of the non-transiting planets from the TTV data. We 
apply our code, {\tt ttvim.f}, to a wide variety of planetary systems to test the uniqueness 
of the TTV inversion problem and its dependence on the precision of TTV observations. We find 
that planetary parameters, including the mass and mutual orbital inclination of planets, 
can be determined from the TTV datasets that should become available in near future. Unlike 
the radial velocity technique, the TTV method can therefore be used to characterize the 
inclination distribution of multi-planet systems. 
\end{abstract}

\keywords{Stars: planetary systems --- celestial mechanics}

\section{Introduction}

In Nesvorn\'y \& Morbidelli (2008) and Nesvorn\'y (2009) (hereafter NM08 and N09) we developed and tested 
a fast inversion method that can be used to characterize planetary systems from the observed Transit Timing 
Variations (TTVs; Agol et al. 2005; Holman \& Murray 2005). See NM08 and N09 for a technical description 
of the method. Here we use this new method to solve the TTV inversion  problem for an arbitrary planetary 
system. The results provide a baseline for studies of real exoplanetary systems for which TTVs will be
detected. Examples of past work that would greatly benefit from the application of the fast inversion 
algorithm discussed here include Steffen \& Agol (2005), Agol \& Steffen (2007), Miller-Ricci et al. (2008) 
and Gibson et al. (2009). 

In \S2, we briefly describe the TTV inversion method. In \S3, we apply it to a case with coplanar 
planetary orbits. Inclined planetary orbits are discussed in \S4. We show, for example, that the 
mutual inclination of planetary orbits can be determined from TTVs. This important parameter, which may be 
used to test planet-migration theories (e.g., Rasio \& Ford 1996, Goldreich \& Sari 2003), is not 
typically available from other existing planet-detection methods.

\section{Inversion Method}

Our TTV inversion method, hereafter TTVIM, has two parts. The first part is a fast algorithm for the 
computation of transit times, $(\delta t_j)_{\rm trial}$, $1\leq j \leq N$, for specified planetary 
parameters. This algorithm is based on perturbation theory (NM08, N09). It calculates the short-period
TTVs as these have been shown to be the most diagnostic (NM08). The long-term effects 
such as the apsidal precession produced by the perturbing planet are more difficult to detect if transit 
observations span only a few years (Miralda-Escud\'e 2002; Heyl \& Gladman 2007). 

The second part of TTVIM is an adaptation of the Downhill Simplex Method (DSM; Press et al. 1992). The 
DSM is used to search for the minima of 
\begin{equation}
\chi^2 = \sum_{j=1}^N \left[  (\delta t_j)_{\rm trial}-(\delta t_j)_{\rm obs}
\over \sigma_j \right]^2\ ,
\label{chi2}
\end{equation} 
where $(\delta t_j)_{\rm trial}$ are the transit times produced by the first part of the algorithm for a 
trial planetary system, $(\delta t_j)_{\rm obs}$ are the observed mid-transit times, and $\sigma_j$ are 
the measurement errors. It is assumed here (as indicated by $\delta$'s) that the period of the transiting 
planet, $P$, has been removed from transit observations. Thus, $(\delta t_j)_{\rm obs} = (t_j)_{\rm obs} - 
C_j P$, where $(t_j)_{\rm obs}$ are the actual transit times and integer $C_j$ denotes the transit 
cycle. 

The best-fit planetary parameters correspond to the global minimum
of $\chi^2$, denoted by $\chi^2_{\rm min}$ in the following. A large number of initial trials must be used 
to assure that the DSM method finds $\chi^2_{\rm min}$. The confidence levels for the normally distributed 
data can be defined as $\Delta \chi^2 = \chi^2 - \chi^2_{\rm min} < (\Delta \chi^2)_{\rm cut}$, 
where the $(\Delta \chi^2)_{\rm cut}$ values are properly chosen for $N$ and the required confidence level
(NM08). 

Here we assume that the mass and orbit of the transiting planet are known from transit and radial velocity 
(RV) measurements as this should be the most common case in practice. If so, $\chi^2$ is a function of seven 
unknown parameters of the perturbing planet, \\$\chi^2 = \chi^2(m_2,a_2,e_2,i_2,\Omega_2,\varpi_2,\lambda_2)$,
where $m_2$ is the mass, $a_2$ semimajor axis, $e_2$ eccentricity, $i_2$ inclination, $\Omega_2$ nodal
longitude, $\varpi_2$ periapse longitude, and $\lambda_2$ the mean orbital phase at $t=0$ (arbitrarily 
defined here to correspond to cycle $C_0$).\footnote{These are the actual parameters used in DSM. The 
boundary at $e_2=0$ does not need a special treatment because $(e_2,\varpi_2)$ is formally equivalent to 
$(-e_2,\varpi_2+\pi/2)$. Similar rules apply to $(i_2,\Omega_2)$ and $(m_2,\lambda_2)$.} 
The parameters of the transiting planet will be denoted by index 1. DSM must therefore search in 
7D space for the global minimum of $\chi^2$. This is not a trivial task because $\chi^2$ often has 
many deep and narrow local minima (Steffen \& Agol 2007). Fortunately, several simplifications 
can be made. 

First, as the amplitude of the short-period TTVs scales linearly with $m_2$, we can calculate the TTV 
profile for the selected $a_2,e_2,i_2,\Omega_2,\varpi_2,\lambda_2$ values and obtain $m_2$ by the linear 
least-square fit. Second, the determination of $\chi^2$ for a new set of the $\Omega_2$, $\varpi_2$ 
and $\lambda_2$ values is computationally 
cheap in the perturbation algorithm, if $\chi^2$ was determined previously for the required $a_2$, $e_2$ 
and $i_2$ values.\footnote{This is because all Fourier terms can be pre-computed for $a_2$, $e_2$ and $i_2$ and 
need only to be assembled with the specific $\Omega_2$, $\varpi_2$ and $\lambda_2$ values. The assembling
procedure itself is computationally inexpensive.} The code can thus efficiently search for the minimum of 
$\chi^2(a_2,e_2,i_2)$ for {\it any} value of $\Omega_2$, $\varpi_2$ and $\lambda_2$. In practice, we use 5 
to 20 values between 0 and 360$^\circ$ to resolve each of these parameters. This effectively reduces the 
number of dimensions to 3. Once the interval of estimated $a_2$, $e_2$ and $i_2$ values is narrowed down, 
the solution can be refined by using the full DSM search in 7D.

The tricky part of TTVIM is the choice of the initial guess in the $(a_2,e_2,i_2)$ space. By trials and errors 
we found that fine (and non-linear) sampling of $a_2$ is generally needed for a successful convergence of the
algorithm. The best results were obtained with uniform sampling in $1/\alpha^2$, 
where $\alpha=a_1/a_2<1$. Parameters $e_2$ and $i_2$ require less care since DSM usually finds the right 
minimum even in the high-$e_2$ and/or high-$i_2$ case if at least one corner of the initial simplex is 
stretched to $e_2>0.2$ and $i_2>30^\circ$. 

With the nominal setup, our TTVIM code ({\tt ttvim.f}) requires about 2 minutes of CPU time\footnote{On a 
single 2.7-GHz Opteron processor.} for a coplanar fit with $i_2=0$ and about 50 minutes for the full 7D fit.
In the absence of measurement errors, the success rate in finding $\chi^2_{\rm min}$ is better than 95\%. 
Thus, {\tt ttvim.f} is a robust code that can reliably solve the TTV inverse problem at a low computational 
cost.  

\section{Results for Coplanar Orbits}

We used a random number generator to define different sets of parameters $m_2$, $a_2$, $e_2$, $i_2$, $\Omega_2$,
$\varpi_2$ and $\lambda_2$. Typically, 1000-2000 different planet parameter sets were used in tests.
In each case, the orbital evolution of the two planets was followed for a fixed timespan, $0<t<T_{\rm int}$, 
with the Bulirsch-Stoer integrator (Press et al. 1992). During this timespan we interpolated for and recorded 
all transit times of the inner planet. This data mimics the real observations, $(\delta t_j)_{\rm obs}$. They 
were used in a blind test where we applied the TTVIM code to each of these cases in an attempt to recover the 
original mass and the orbital elements of the non-transiting planet.

We start by discussing the case with star's mass $m_0 = M_{\rm Sun}$, where $M_{\rm Sun}=2\times10^{33}$ g
is the mass of the Sun, $m_1=10^{-3}\ m_0$, $a_1 = 0.1$ AU, $e_1=i_1=0$ and $N=100$ consecutive 
transits. Since NM08 and N09 showed that the behavior of the inversion method is insensitive to $m_1$, 
we will not test different $m_1$ values in this work. To distinguish between the issues related to
the intrinsic limitations of {\tt ttvim.f} and those arising from the finite precision of the real 
measurements, we first discuss an idealized case with zero measurement noise. 

\begin{figure}
\epsscale{0.7}
\plotone{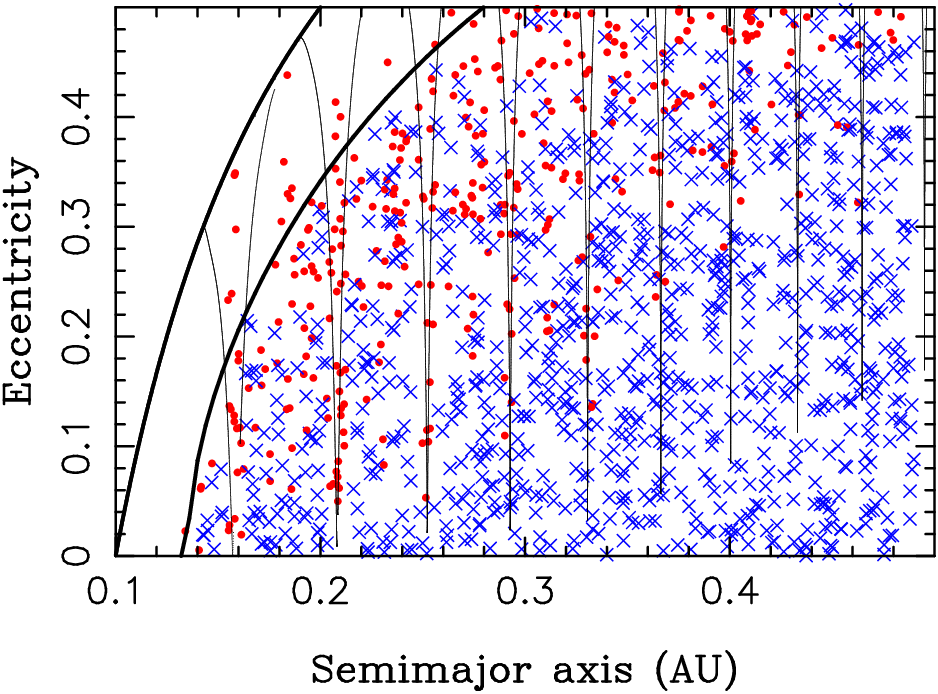}
\caption{TTVIM code results for planetary systems with $m_0=M_{\rm Sun}$, $m_1 = 10^{-3} m_0$, $a_1=0.1$ AU,
$e_1=0$ and $i_2=0$. Planetary parameters for which the TTVIM code converged to the correct solution
were denoted by blue $\times$'s. Incorrect solutions were denoted by red dots. We defined the correct solution 
as having $|m_2 - m_2^*|/m_2<0.5$, $|a_2 - a_2^*|/a_2<0.05$ and $|e_2 - e_2^*|<0.05$, where $m_2$, $a_2$ 
and $e_2$ are the original planetary parameters for which the TTV signal was computed by $N$-body 
integration, and $m_2^*$, $a_2^*$ and $e_2^*$ are the values determined by the TTVIM code. In the 
majority of cases corresponding to correct solutions, the TTVIM code determined the original orbital  
parameters with a better than 2\% precision and mass with a better than 20\% precision.
 The two bold solid lines show the planet-crossing (upper) and 
Hill-stability limits (lower). We also show the location of the principal mean motion resonances between 
the planets (e.g., 2:1 at $a_2=0.16$ AU). 
There are two lines per resonance corresponding to the left and right separatrices of 
resonant motion. The V-shaped profiles are characteristic for mean motion resonances that become wider 
with eccentricity.}
\label{case1}
\end{figure}

For coplanar orbits and $\sigma_j=0$, the TTVIM code finds the correct planetary parameters with
a high rate of success (Fig. \ref{case1}). The typical precision in the successful cases is 
$|m_2 - m_2^*|/m_2<0.2$, $|a_2 - a_2^*|/a_2<0.02$, $|e_2 - e_2^*|<0.02$, 
$|\varpi_2 - \varpi_2^*|<10^\circ$ and $|\lambda_2 - \lambda_2^*|<10^\circ$, where the asterisk denotes 
the values determined by the TTVIM code.\footnote{Except for very small values of $e_2$ for which the
errors in $\varpi_2$ can be large.} This is very satisfactory. In the absence of measurement errors, 
the result of the TTVIM code illustrated in Fig. \ref{case1} with $m_2 = 10^{-4} m_0$ is insensitive to 
the actual value of $m_2$. 

The main failure mode of the TTVIM code occurs near mean motion resonances between planets, because
resonant perturbations are not (yet) taken into account in TTVIM. While the resonant signal can improve 
our chances of the TTV detection for (near-)resonant planets, it seems less useful in helping us 
estimate the mass and orbit of the planetary companion. Specifically, the amplitude and period of the 
resonant signal can be fit by a number of different planetary setups corresponding to different resonances. 
Thus, without an apriori knowledge of the mean motion resonance that is responsible for the observed 
behavior, the inversion problem from resonant frequencies alone is strongly degenerate.

Fortunately, the short-period TTVs underlying the resonant signal can still be used to determine the
planetary parameters without much ambiguity. As shown in NM08, probably the best strategy is to isolate 
short-period frequencies in the signal by Fourier filter and apply the inversion method to the filtered 
signal. The application of this procedure is straightforward in individual cases (see NM08), where the 
resonant period, and thus the appropriate frequency cutoff, can be estimated from $(t_j)_{\rm obs}$. 
We verified that this procedure works quite well in $>$75\% of cases shown in Fig. \ref{case1} in which 
the resonant variations are an issue.

The remaining $<$25\% unsuccessful cases (representing $<$5\% overall) correspond to the very large 
values of $e_2$ for which the Laplacian expansion of the perturbing function in TTVIM is not convergent
(NM08), and/or planetary configurations that are not Hill stable. Direct $N$-body integrations can
be used to address the TTV inversion problem in the very-high-eccentricity domain, but the CPU cost of 
these tests is likely to be substantial and lies beyond the scope of this letter.

\begin{figure}
\epsscale{0.5}
\plotone{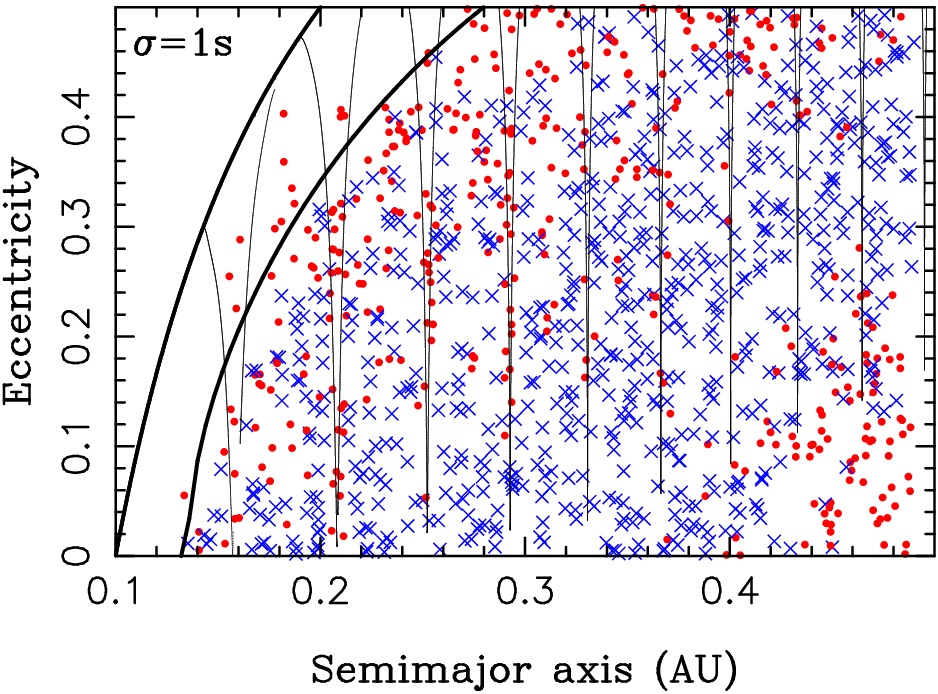}\hspace*{5.mm}
\plotone{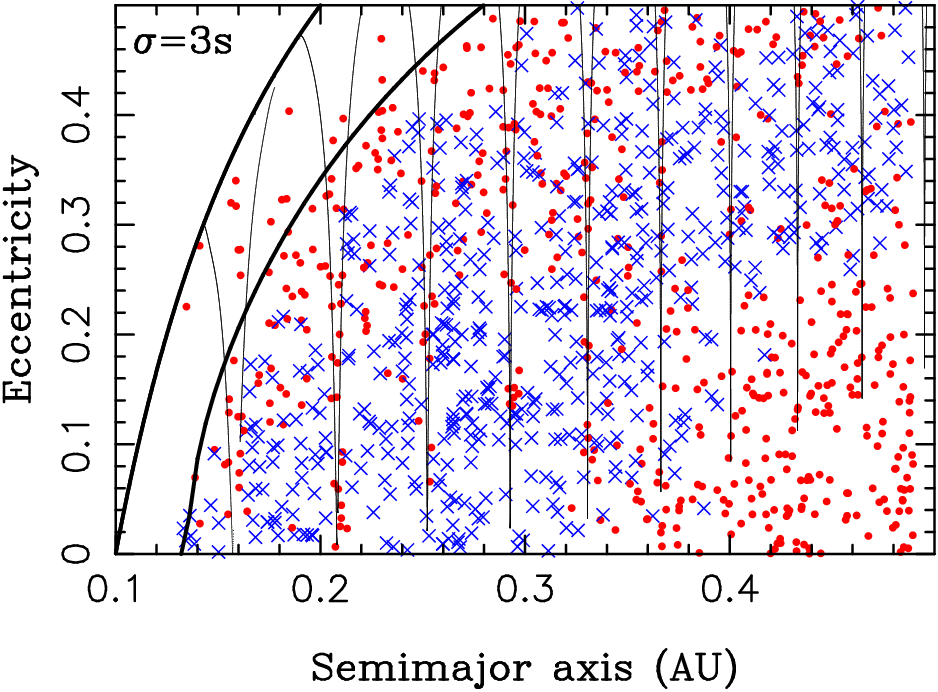}\vspace*{5.mm}
\plotone{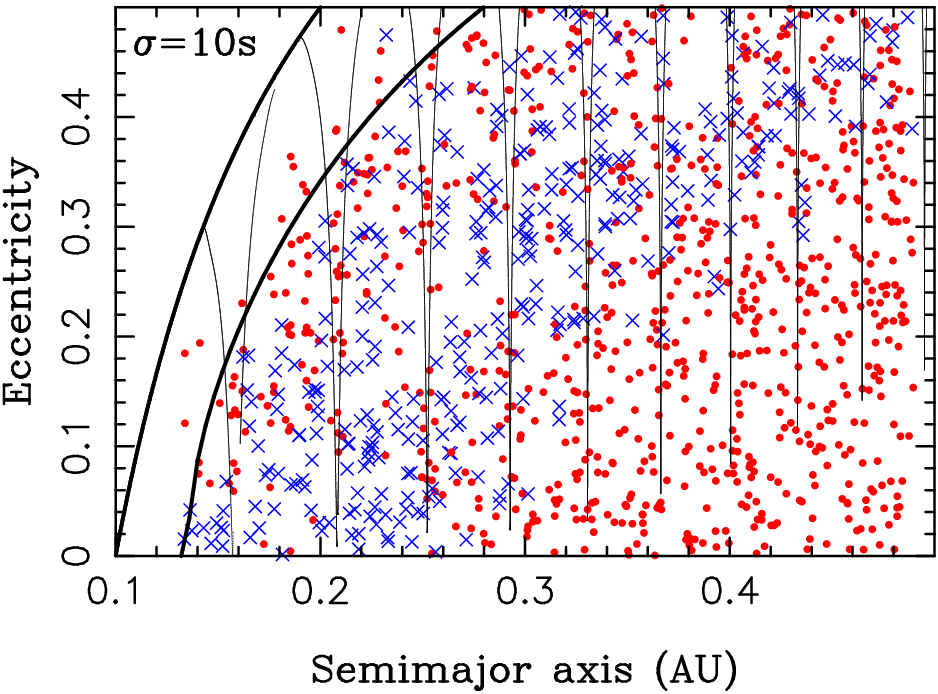}\hspace*{5.mm}
\plotone{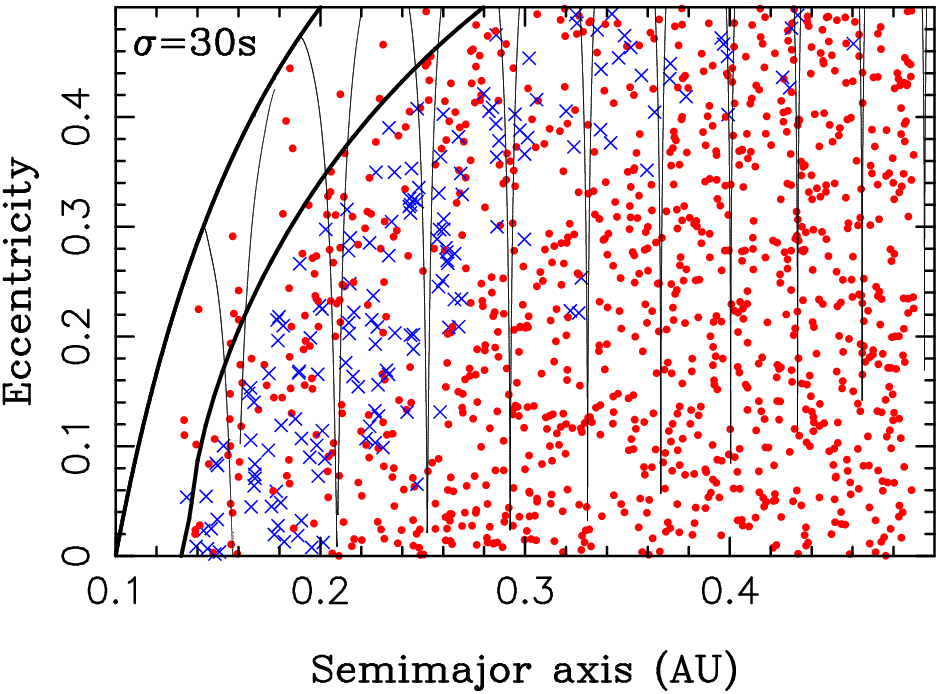}
\caption{TTVIM code results for a planet with $m_2=10^{-4} m_0$ and different levels of the measurement 
error, $\sigma$. See caption of Fig. \ref{case1} for a description of different lines and symbols.}
\label{case2}
\end{figure}

The measurement errors have a profound effect on the uniqueness of the inverse problem (Fig. \ref{case2}). 
For $m_2=10^{-4}\ m_0$, $N=100$ and $\sigma_j=\sigma=3$ s, corresponding to the Kepler-like precision of 
timing measurements for a Sun-like star with a 2 Neptune-mass planet, unique determination of planetary
parameters can be achieved for most stable systems with $q_2 = a_2(1-e_2)<3.3 a_1$, while for $\sigma_j=\sigma=10$ 
s (Corot-like precision), it is required that $q_2 < 2.6 a_1$. These limits approximately correspond to
the planetary parameters for which the amplitude of the short-period TTVs is comparable to $\sigma$. 

\begin{figure}
\epsscale{0.5}
\plotone{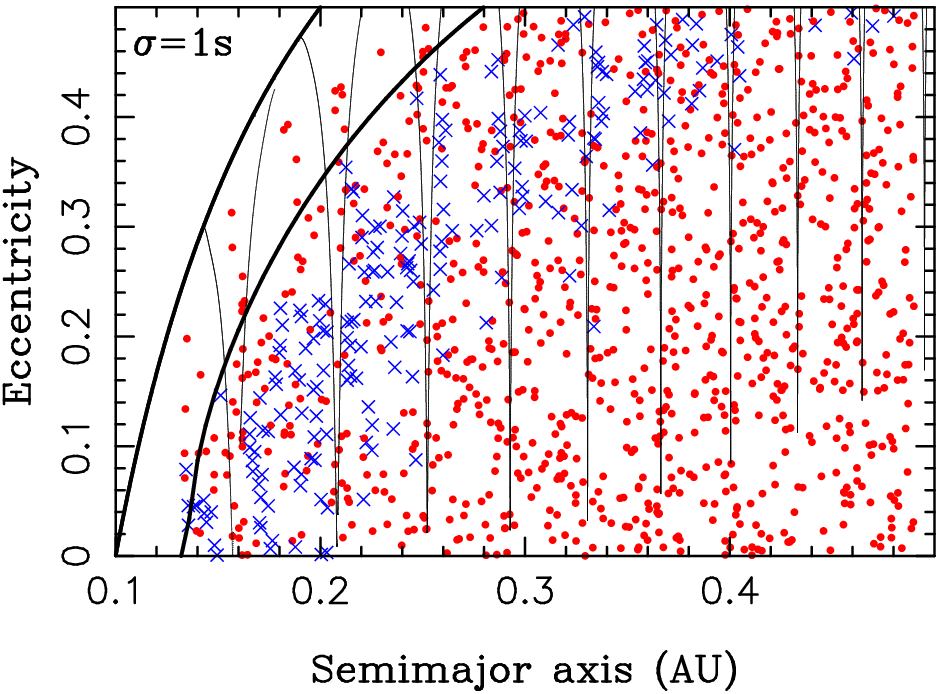}\hspace*{5.mm}
\plotone{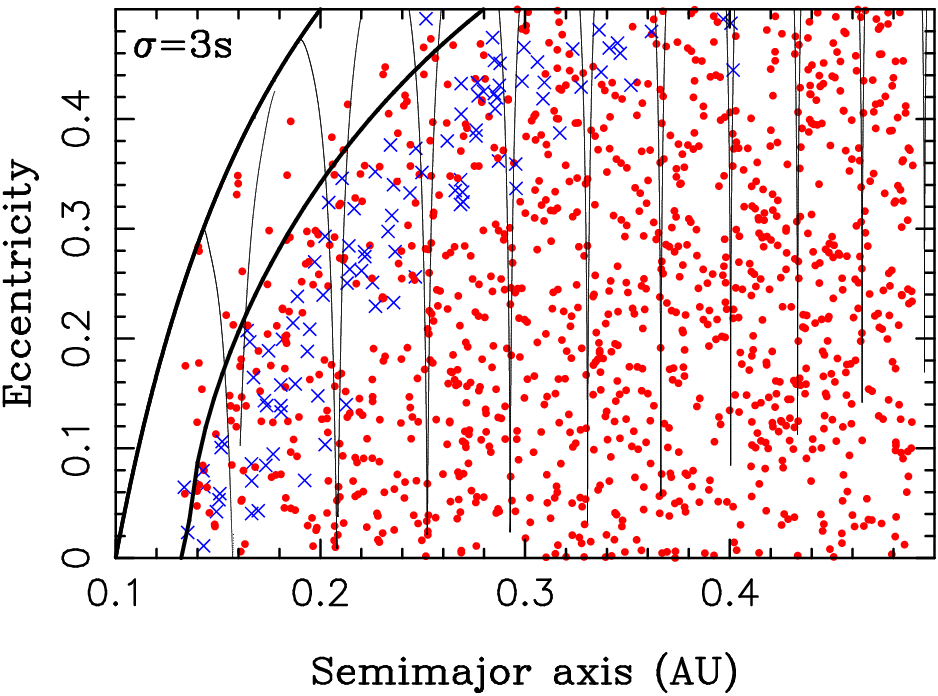}
\caption{TTVIM code results for an Earth-mass planet ($m_2=3\times10^{-6} M_{\rm Sun}$) and two different 
levels of the measurement error, $\sigma$. See caption of Fig. \ref{case1} for a description of different 
lines and symbols. With $\sigma>3$ s, the TTVIM code can only characterize the Earth-mass planets 
with very specific orbits.}
\label{case2e}
\end{figure}

Figure \ref{case2e} shows the result of the TTVIM code for an Earth-mass planet. The region of parameter
space in which unique determination can be achieved from TTVs is relatively small even with $\sigma=1$ s. 
Thus, an Earth-mass planet detection {\it and} characterization of its orbit will require a rather 
fortuitous  setup of the planetary system, in which $(q_2-a_1)/a_1 \lesssim 2$ (for external perturber).

\section{Results for Inclined Planetary Orbits}

\begin{figure}
\epsscale{0.5}
\plotone{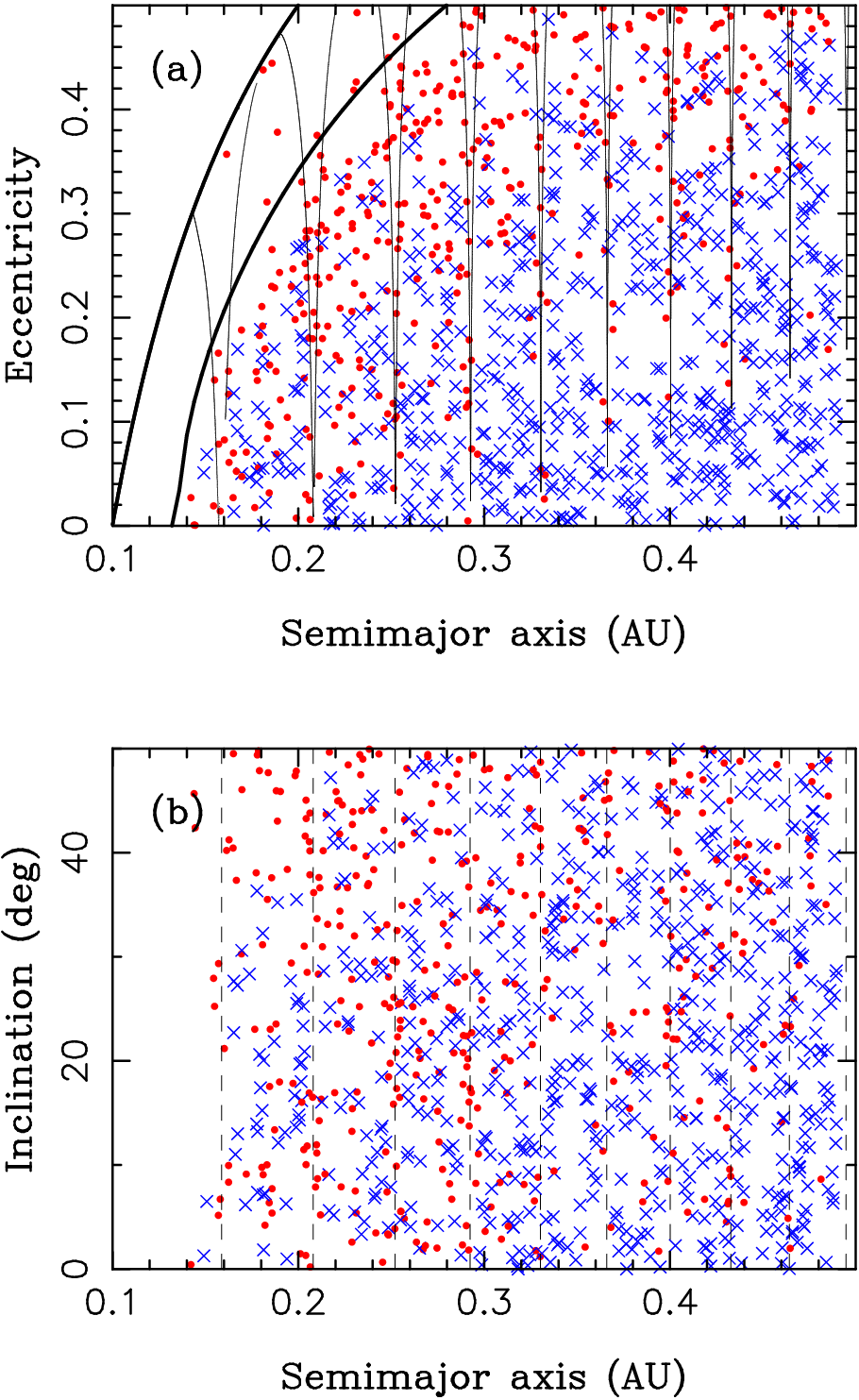}
\caption{The same as Fig. \ref{case1} but with $i_2 \neq 0$. Most correct solutions (blue $\times$'s)
have $|a_2 - a_2^*|/a_2<0.015$, $|e_2 - e_2^*|<0.02$, $|i_2 - i_2^*|<2^\circ$ and $|m_2 - m_2^*|/m_2<0.2$.
In (b), the dashed lines denote the libration centers of the principal mean motion resonances.}
\label{case3}
\end{figure}

We applied the TTVIM code to mock planetary systems with $0<i_2<50^\circ$. As in \S3, we assumed that
$m_0 = M_{\rm Sun}$, $m_1=10^{-3}\ m_0$, $a_1 = 0.1$ AU, $e_1=0$ and used $N=100$ consecutive transits.
Figure \ref{case3} shows the result for $\sigma_j=\sigma=0$. The $(a,e)$ plot does not differ much
from the coplanar case although it may be noted that the quality of fits slightly degraded for
the large $e_2$ values. This is probably related to the convergence problems of the perturbation 
algorithm in TTVIM. A precise $N$-body integrator should perform better for high $e_2$, although
it has yet to be shown that an $N$-body integrator can be applied to the inclined inverse problem in 
practice due to the large CPU cost.

\begin{figure}
\epsscale{0.5}
\plotone{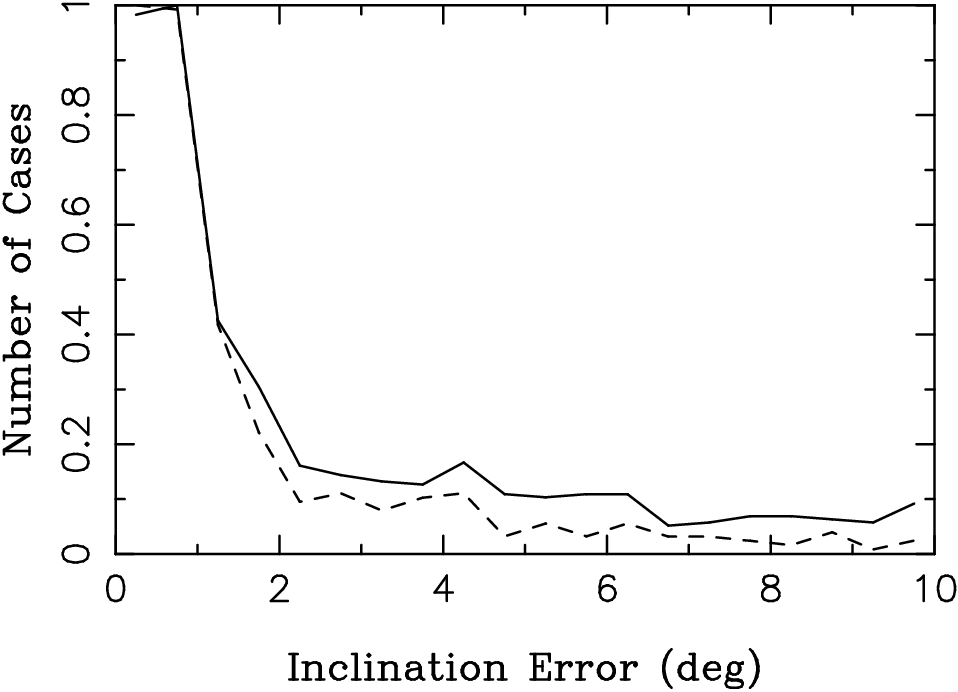}\hspace*{5.mm}
\plotone{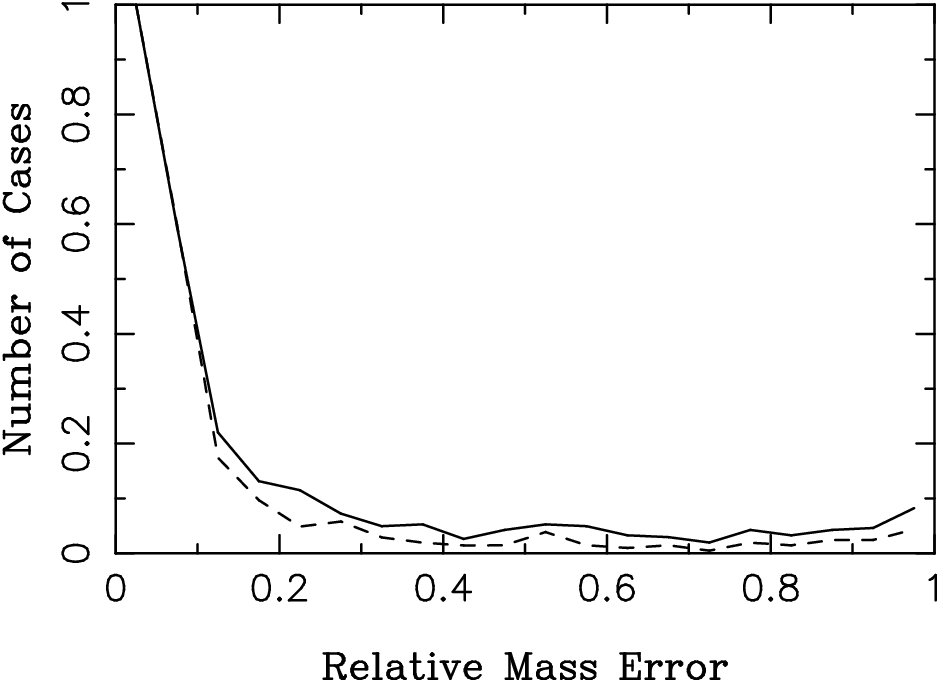}
\caption{Distribution of TTVIM errors in $i_2$ (left) and $m_2$ (right) for the case shown in Fig. 
\ref{case3}. We show the total distribution (solid line) and the one for $q_2>0.25$ AU (dashed). 
In the later case, the erroneous determinations with $|i_2 - i_2^*|>5^\circ$ are reduced because 
the algorithm does not need to deal with the difficult case when $q_2 \sim a_1$. Most cases 
correspond to $|m_2 - m_2^*|/m_2<0.2$ (i.e., $<$20\% precision of mass determination) and 
$|i_2 - i_2^*|<2^\circ$.}
\label{istat}
\end{figure}

Probably the most exciting result obtained in this work is that it was possible to determine the mutual 
inclination of planets for most planetary systems (Fig. \ref{case3}b). Unlike the radial velocity technique, 
the TTV method can therefore be used to characterize the inclination distribution of multi-planet systems. 
Figure \ref{istat} shows the detailed statistic of TTVIM errors in $i_2$. In most
cases, $|i_2 - i_2^*|<2^\circ$. The tail of larger $|i_2 - i_2^*|$ values corresponds to the 
high-eccentricity cases. If the statistic is limited to $q_2=a_2(1-e_2)>0.25$ AU (Fig. \ref{istat}a;
dashed line), the fraction of successful cases with $|i_2 - i_2^*|<2^\circ$ increases 
to $>$90\%. In the succesful cases, orbital angles $\Omega_2$, $\varpi_2$ and $\lambda_2$ are generally
correctly determined to within a better than 5$^\circ$ precision.

We also studied how the uniqueness of the inclined inverse problem is affected observational
errors. The trends seen in these tests are very similar to those described in \S3. Namely, the instrumental 
noise sets an upper limit on $q_2$ beyond which the determination of planetary parameters from TTVs 
is ambiguous. Again, we see that these limits approximately correspond to the planetary parameters for which 
the amplitude of the short-period TTVs is comparable to $\sigma$. The results in N08 and 
Payne et al. (2009) can therefore be used to estimate whether (or not) a unique characterization of 
the specific inclined planetary system may be achieved from TTV observations with given $\sigma$.

We find that TTVs obtained in the coplanar case represent a good approximation of the TTVs for planetary 
orbits with $i_2<20^\circ$. The planar version of the TTVIM algorithm can therefore be used in these 
cases to estimate the $a_2$ and $e_2$ values of the perturbing planet. This helps to narrow the range of initial 
guesses for the 7D fit and represents a factor of $\sim$20 speed up of the inversion. The full 7D 
algorithm needs to be used for $i_2>20^\circ$.

\section{Conclusions}

The method developed here can be used to analyze TTVs found for any of the potentially hundreds of
planets expected to be discovered by {\it Kepler} (Beatty \& Gaudi 2008). {\it Kepler} should be able
to detect transit timing variations of only a few seconds (Holman \& Murray 2005), which should easily 
exist in many systems, extrapolating from the radial velocity planets (Agol et al. 2005, Fabrycky 2008).

Perhaps the most interesting result that comes out of this work is that the shape of the TTV signal is
generally sensitive to the orbital inclination of the non-transiting planetary companion. Thus, the TTV
method can provide means of determining mutual inclinations in systems in which at least one planet is
transiting. This parameter cannot be determined by other planet-detection methods. 

TTVIM algorithm can be easily extended to incorporate uncertainties in the transiting planet's parameters. 
This can be done by sampling dimensions that correspond to the additional parameters. For example, in N09 
we extended the NM08 method to the case with $e_1\neq0$. This may be especially relevant to the transiting 
planets that will be discovered by Kepler because these planets are expected to have wider orbits, which 
are less susceptible to the circularizing effects of tides. The low CPU cost of the TTVIM algorithm is the 
key element which will make such studies possible. 

\acknowledgements

This work was supported by the NSF AAG program.

\clearpage

\clearpage

\clearpage

\clearpage

\clearpage

\end{document}